# Crossed-Time Delay Neural Network for Speaker Recognition

Liang Chen[a], Yanchun Liang[b], Xiaohu Shi[a], You Zhou[a], Chunguo Wu[a, b*]

[a]Key Laboratory of Symbolic Computation and Knowledge Engineering of Ministry of Education, College of Computer Science and Technology, Jilin University, Changchun, 130012, PR China
[b]Zhuhai Laboratory of Key Laboratory of Symbol Computation and Knowledge Engineering of Ministry of Education, School of Computer, Zhuhai College of Jilin University, Zhuhai, 519041, PR China

*Corresponding E-mail: wucg@jlu.edu.cn*

**Abstract.** Time Delay Neural Network (TDNN) is a well-performing structure for deep neural network-based speaker recognition systems. In this paper we introduce a novel structure, named Crossed-Time Delay Neural Network (CTDNN) to enhance the performance of current TDNN for speaker recognition. Inspired by the multi-filters setting of convolution layers from convolution neural networks, we set multiple time delay units with different context size at the bottom layer and construct a multilayer parallel network. The proposed CTDNN gives significant improvements over original TDNN on both speaker verification and identification tasks. It outperforms in VoxCeleb1 dataset in verification experiment with a 2.6% absolute Equal Error Rate improvement. In few shots condition, CTDNN reaches 90.4% identification accuracy, which doubles the identification accuracy of original TDNN. We also compare the proposed CTDNN with another new variant of TDNN, Factorized-TDNN, which shows that our model has a 36% absolute identification accuracy improvement under few shots condition. Moreover, the proposed CTDNN can handle training with a larger batch more efficiently and hence, utilize calculation resources economically. The code of the new model is released at https://github.com/chenlliang/CTDNN

## 1    Introduction

The Speaker recognition system verifies or identifies a speaker's identity based on speaker's voice. It can be divided into speaker verification and speaker identification, where speaker verification aims to verify whether an utterance corresponds to a given identity and speaker identification aims to identify a speech from all enrolled speakers. According to the different testing scenario, speaker recognition can also be categorized into closed-set or open-set settings. For closed-set scenario, all testing identities are enrolled in the training set, therefore it can be regarded as a classification problem. For open-set scenario, the testing identities are not previously seen in the training set, which is closer to real world application since new identities will be added to the system continually. To address that problem, each utterance must be mapped into an embedding space where cosine similarity is used to evaluate whether two utterances correspond to one same identity. The current studies mainly focus on the open-set speaker recognition problem.

Recently, deep neural network has been widely applied to learning speakers' embedding through the learning process of classification such as x-vectors and have shown great priority in performance [1] than traditional statistical models such as HMM-GMM [2] and i-vectors. Time delay layer is an important component among DNN-based models due to its ability to capture feature from sequent audio data.

Time Delay Neural Network was regarded as the ancestor of convolution neural network [1]. It is effective in capturing features from long range temporal contexts and is widely used in speech related field such as speaker recognition system automatic speech recognition and speech synthesis [3]. The TDNN architecture, shown in Figure 1, uses a modular and incremental method to create larger networks from sub-components [4]. The time delay architecture can be regarded as a convolution on sequence data where a 1-dimension filter scans through the input sequence and generate an output at each step with the strategy of weight-sharing. Many related works have focused on TDNN, such as TDNN-LSTM[5], TDNN-BLSTM[6], CNN-LSTM-TDNN [7] and FTDNN [8]. References [5] [6] [7] focus on combining TDNN with different components to construct better model and reference [8] proposed a variant of TDNN through low-rank matrix factorization and skip connection to overcome gradient explosion problem for TDNN-based network structures.

We propose the crossed-time delay neural network as a variant of TDNN, named CTDNN. The multiple-filters mechanism of a convolution layer from CNN inspires us to set different time delay units at the bottom layer of the network. In CNN, each filter with different parameters in the same convolution layer captures different characteristics of the input by generating different feature maps, which ultimately helps to classify the input image. In the original TDNN, there is only one filter per layer, which restricts the model's feature extraction and generalization ability according to our analysis and experiments. The proposed structure has three main advantages:
• The time delay units with different context size in the same layer help to extract more heterogeneous features.
• The structure is wider, but not deeper, which avoids gradient explosion and vanishing problem arising occasionally in the training process and guarantees the generalization ability.
• Our model works well with large batch, compared to FTDNN, which enables it to utilize calculation resources in a more efficient way without alternating the batches frequently.

## 2  Baseline Models

The network architecture of our speaker recognition baseline systems is the same as the original x-vector system in [1] and the improved architecture FTDNN in [8].

The TDNN architecture shown in Figure 1 is applied in x-vector system. Initial transforms are learnt on narrow contexts and the deeper layers process the hidden activations from a wider temporal context. Hence the higher layers have the ability to learn wider temporal relationships [1]. The time delay architecture can be regarded as a one-dimension convolution on sequence data where a 1-d filter scans through the input sequence by the strategy of weight-sharing. The time delay layers is followed by the statistical pooling(SP) layer which computes the statistical feature by computing the mean and variance. Fully connected layers and SoftMax are posed after the SP layer to project the sequence into speaker's identity. During back-propagation, the lower time delay layers are updated by a gradient accumulated over all the time steps of the input sequence. Thus, the lower layers of the network are forced to learn translation invariant feature transforms [1].

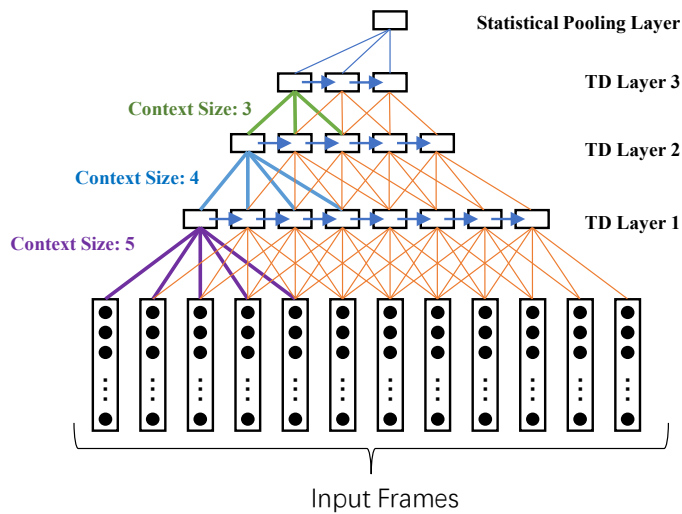

Figure 1. An example for the original TDNN. Time Delay (TD) Layer 1,2 and 3 each has a context size of 5,4 and 3

The FTDNN is a factored form of TDNN which is structurally similar to TDNN, whose layers have been compressed via singular value decomposition to reduce the number of parameters, and uses shortcut connection [9] and highway connections in order to avoid gradient diffusion problems in deeper neural network.

## 3    Crossed-Time Delay Neural Network

The proposed CTDNN is shown in Figure 2, which is a wide and shallow structure rather than a narrow and deep structure. It combines the Crossed-Time delay layers and the statistical pooling layers.

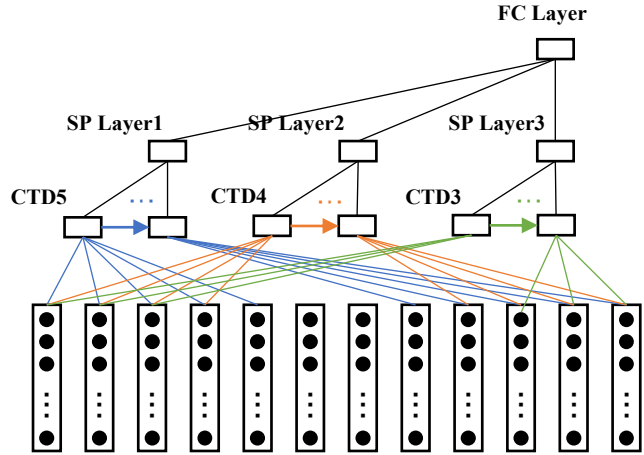

Figure 2. The structure of CTDNN. Three parallel Time Delay units as CTD 5,4 and 3 are marked.

### 3.1    Crossed-Time Delay Layer

We set different time delay units in the bottom layer to directly extract features from the input sequence. Each unit has a different context frame size, which means they take a different number of frames of MFCC feature as input. In Figure 2, the context frame sizes are 5, 4 and 3 marked as Crossed-Time Delay (CTD) unit 5, CTD unit 4 and CTD unit 3. Each unit scans the input sequence separately and output a fixed size vector at each step till the end of the sequence. In other words, we can regard the different time delay units as different filters and each of them take the sequence as input and outputs different feature maps. The CTDNN layers can also be stacked vertically to form a deeper hierarchy structure, in this case, each feature map should be allocated a new time delay unit.

It seems against the consensus that the deep and narrow network is better than the wide and shallow one as discussed in [10]. However, the extension of the layer width is not to simply add more neurons and connections, but to extract features at different frequencies or paces. We exploit the strength of the structure from two perspectives including the heterogeneous feature extraction and the more feasible training process.

**Heterogeneous Feature Extraction**

Using crossed-time delay units can extract more heterogeneous feature than that of a TDNN. Since the raw audio is viewed as short-time stationary signal, it has to be framed to short-time pieces at a fixed frequency to further analyze the audio and extract other features like MFCC. In original TDNN models, the time delay units are stacked vertically, and each unit has fixed reception field and parameters within connections. This single-line structure has the bottom layer to domain the feature extraction capacity, which limits the generalization ability of the model.

Take the model shown in Figure 1 as an example. The bottom layer has a context size of 5, so it takes in 5 frames of MFCC feature at a time. The second layer has an input size of 4, and it takes in four features from the bottom layer as input, which enlarges its context size to 8 dues to the tree-like vertical structure. However, the second layer does not actually take input from a context size of 8 but the linear combination of 4 short sequences at the size of 5. So does the deeper layers. So, the key of the model is up to the bottom layer. With a fixed set of parameters and context size, the feature it gets is homogeneous since there are features that range more or less than 5 frames because of the short-time stationary property of audio signal and those features cannot be captured by one fixed-context-size time delay unit.

As shown in Figure 2, we set 3 time-delay units each with a different context size at the bottom layer. During back-propagation, due to the different context size, the lower layers of the network are updated by a gradient accumulated over different time steps of the input temporal context. Hence, the lower layers of the network are forced to learn different feature transforms, which enlarges the feature extraction capacity of the model.

**Training Convenience**

Shallow networks are more feasible to train and converge, especially on small datasets. Training might suffer from gradients vanishing or exploding problems during the process of back-propagation in deep neural network. The literature [10] found that relatively small network sizes have obvious computational advantages when training on small dataset. We leverage the depth and width of CTDNN in our experiments and find that building two CTDNN layers can outperform 5 normal TDNN layers in both common and few-shots learning tasks.

### 3.2    Statistical Concatenation

Since the context size of time delay unit differs in the bottom layer, the output of the units will have different length. Instead of doing statistical pooling on all the output in one time, we compute the mean and standard deviation for each time delay unit's output and concatenate the results parallel before the fully connected layer.

# 4    Experiments

We conducted our experiments on the open VoxCeleb1[12] dataset and VCC2016[13] dataset to test the models' performance under large and few samples condition. To be more specific, an open-set text-independent verification experiment was performed with VoxCeleb1 dataset and a close-set text-independent classification experiment was done with VCC2016 dataset since it has limited number of speakers.

Table 1．Dataset details. Average Length is the average number of frames after preprocessing in each dataset.

| Dataset Name | Num Speakers | Total Utterances | Average Length |
|---|---|---|---|
| VoxCeleb1 | 1251 | 14425 | 791 |
| VCC2016 | 10 | 1134 | 270 |
| VCC2016(Mini) | 10 | 162 | 255 |

## 4.1    Preprocessing

For VoxCeleb1 the acoustic features were 30-dimensional MFCC features extracted every 10ms and the frame size for short-time Fourier transform (STFT) was 25ms. And for VCC2016 the acoustic features were 13-dimensional MFCC features and others are the same with VoxCeleb1. In order to obtain the same length inputs, we duplicate the short-length input and cut off the extra length to make all the input at the same length (1000 frames for Voxceleb1 and 300 for VCC2016). No more enhancing or aligning methods were implemented. Our model was implemented with PyTorch.

## 4.2    Model Configuration

Table 2. Model Settings

|  | TDNN | FTDNN | CTDNN |
|---|---|---|---|
| 1 | TD [-2,2] | TD [-2,2] | CTD [-4,4]; CTD [-2,2]; CTD [-1,1] |
| 2 | TD [-1,2] | FTD Layer | CTD [-1,1] *3 |
| 3 | TD [-3,3] | FTD Layer | SC |
| 4 | TD [7,2] | FTD Layer | FC |
| 5 | SP | FTD Layer | FC |
| 6 | FC | FTD Layer | SoftMax |
| 7 | FC | FTD Layer |  |
| 8 | SoftMax | FTD Layer |  |
| 9 |  | FTD Layer |  |
| 10 |  | FC |  |
| 11 |  | SP |  |
| 12 |  | FC |  |
| 13 |  | FC |  |
| 14 |  | SoftMax |  |

In Table 2, FC stands for the Fully Connected layer, SP for Statistical Pooling Layer and SC stands for Statistical Concatenation. We construct the TDNN and FTDNN structure the same as [1] and [8]. TDNN structure combines of 4 TDNN layers. FTDNN has up to 14 layers, i.e., the deepest structure in our experiments. All the Time Delay layers in the three models have batch normalized input and are activated by ReLU. Dropout and skip-connection policies are only involved in FTDNN not in TDNN and CTDNN. To be recognized, the proposed CTDNN has a wider and shallower structure. We set 3 time-delay units in the first and second layer.

### 4.3    Training Parameters Settings

We used cross entropy as the loss function. Adam optimizer was used and the training batch size was 128 in VCC2016 and 50 in VoxCeleb1. The learning rate was fixed to 0.0001 for CTDNN and TDNN and 0.001 for FTDNN. Early Stopping was used to prevent overfitting.

### 4.4    Embedding Extraction and Verification

Embeddings were extracted after SP layer for TDNN and after SC Layer for CTDNN. Linear Discriminative Analysis was applied to reduce the embeddings' dimensions as well as alleviate the influence due to channel differences. We reduced the dimensions to 400 for CTDNN and 250 for TDNN since the CTDNN embedding's size is larger. We train the LDA model on the training data and apply it to the testing data to evaluate the performance of the systems. Cosine similarity was applied to all test cases in the testing part of VoxCeleb1 dataset to compute the similarity between two utterance and get the EER of the system.

## 5    Results

### 5.1    VoxCeleb1

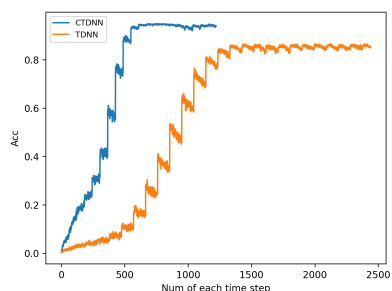

Figure 3. Learning Curve of VoxCeleb1 experiment. Each time step equals to 10 batches. The accuracy was tested on training data.

Figure 3 shows the update of accuracy on training data during the classification training process. The CTDNN structure converged sooner than the original TDNN and reach a higher training accuracy. Table 3 shows the result of TDNN's and CTDNN's identification performance on VoxCeleb1 dataset. CTDNN outperformed the TDNN structure by 30% improvement on EER.

Table 3. EER and Converging Epochs on VoxCeleb1 Dataset

| Structure | EER | Epochs |
|-----------|--------|--------|
| TDNN | 0.054 | 31 |
| CTDNN | **0.0382** | **12** |

## 5.2 VCC2016

Table 4. *Best Top1 Test Accuracy*

| Structure | VCC2016 | VCC2016(Mini) |
|-----------|---------|---------------|
| TDNN | 0.778 | 0.448 |
| CTDNN | **0.992** | **0.904** |
| FTDNN 32 | 0.965 | 0.662 |
| FTDNN 128 | 0.608 | 0.681 |

Table 4 shows the results on two experiments. In both experiments, our CTDNN outperforms the other structures, especially in few samples learning in which the accuracy is more than 2 times of the original TDNN. Moreover, the experiments show that the performance of FTDNN gets worse when batch size grows and can't converge with the batch size of 128. We then tune the batch size and find 32 is the most proper setting for the first experiment. However, FTDNN can't work well with any batch size compared to CTDNN in few samples condition.

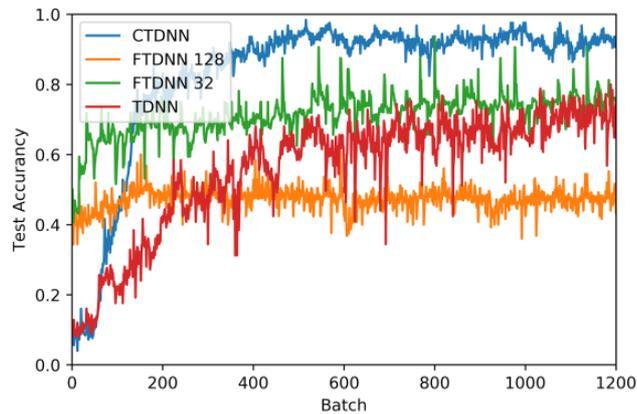

Figure 4. Learning Curve of VCC2016 Experiment

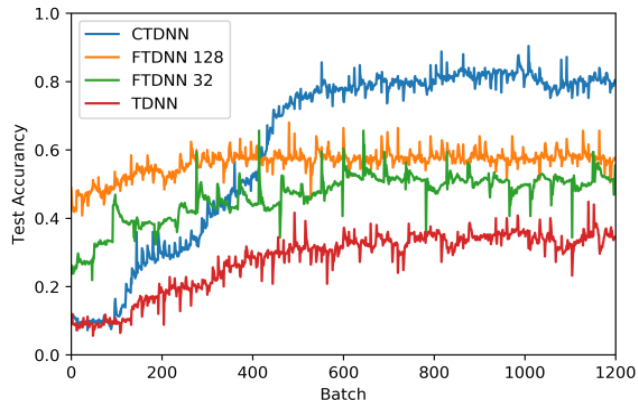

Figure 5. Learning Curve of VCC2016(Minus) Experiment

The reason for FTDNN can't do well with large batch size is actually a general problem as discussed in [11]. There is still no consensus on how to tune batch size for different models. Different models might have different best batch size on different tasks. A large batch can significantly speed up the training while might suffer from loss in accuracy compared with small batch. From that perspective, the fact that CTDNN can achieve higher accuracy with large batch size also suggests that it can take full advantage of the GPU resources and speed up the training process.

Figure 4 and 5 shows the curves of test accuracy during training. It can be seen that CTDNN comes to convergence with higher accuracy than other models in both experiments.

## 6    Conclusion

In this work, we analyzed and examined the performance the new structure CTDNN on both speaker identification and verification tasks. Our analysis suggests that the crossed-time delay units can extract heterogeneous features therefore achieve better feature extraction ability. And our two experiments proved our analysis and showed the of large batch capacity of CTDNN.

TDNN was once the precursor of convolution neural network, now we apply the characteristics from CNN to improve TDNN and gain improvements. In the future we will explore more application of CTDNN such as using it to improve different TDNN based model and combine it with embedding extraction system like x-vector to find out its effect on speaker embeddings. Furthermore, there are still to explore about the relation between CTD units' parameters and feature capturing ability such as the number and the size of each CTD unit.

# 7    Acknowledgements

This work is supported by the National Natural Science Foundation of China (61772227, 61876069, 61876207 and 61972174), the Key Development Project of Jilin Province (20180201045GX and 20180201067GX), the Jilin Natural Science Foundation (20200201163JC), the Guangdong Key-Project for Applied Fundamental Research (2018KZDXM076), and the Guangdong International Cooperation of Science and Technology (2020A0505100018).